\documentclass[aps,prd,twocolumn,preprintnumbers, groupedaddress,nofootinbib,amssymb,notitlepage,eqsecnum]{revtex4-2}
\usepackage{bm}
\usepackage{amsmath,amsthm,amssymb}
\usepackage{amsfonts}
\usepackage{hyperref}
\usepackage{graphicx}

\allowdisplaybreaks[1]

\newcommand{\be}{\begin{equation}}
\newcommand{\ee}{\end{equation}}
\newcommand{\ba}{\begin{eqnarray}}
\newcommand{\ea}{\end{eqnarray}}
\newcommand{\Mpl}{M_{\rm Pl}}

\begin{document}

\preprint{WUCG-25-09}

\title{Crossing the phantom divide in scalar-tensor 
and vector-tensor theories}

\author{Shinji Tsujikawa}

\affiliation{
Department of Physics, Waseda University, 3-4-1 Okubo, 
Shinjuku, Tokyo 169-8555, Japan}
 
\date{\today}

\begin{abstract}

DESI observations of baryon acoustic oscillations (BAOs), combined with cosmic microwave background (CMB) and type-Ia supernova (SN~Ia) data, suggest that the dark energy equation of state $w_{\rm DE}$ crosses the phantom divide from $w_{\rm DE} < -1$ to $w_{\rm DE} > -1$ at low redshifts. In shift-symmetric Horndeski and generalized Proca theories with luminal 
gravitational-wave speed and no direct couplings to dark matter, 
we show that such a  
phantom-divide crossing is generically difficult without theoretical pathologies. 
Breaking the shift symmetry in Horndeski theories allows
this transition. We construct an explicit model with broken shift symmetry, in which 
the scalar field has a potential in addition to a Galileon self-interaction 
and a quadratic kinetic term.
This model realizes the desired phantom-divide crossing at low redshifts 
without introducing ghosts and Laplacian instabilities.

\end{abstract}


\maketitle

\section{Introduction}
\label{Intro}

Recent DESI observations of BAOs challenge 
a cosmological constant and support dynamical dark energy (DE) models, 
in which the DE equation of state $w_{\rm DE}$ evolves with time \cite{DESI:2024mwx,DESI:2024aqx}.
When the DESI data release 2 measurements \cite{DESI:2025zgx} 
are combined with CMB observations \cite{Planck:2018vyg}, 
there is a $3.1\sigma$ preference for dynamical DE based on 
the Chevallier-Polarski-Linder parametrization of 
$w_{\rm DE}$ \cite{Chevallier:2000qy,Linder:2002et}.
This trend persists when SN~Ia data are included, 
showing a preference for dynamical DE with a significance between 
$2.8\sigma$ and $4.2\sigma$, depending on the dataset 
combination \cite{DESI:2025fii}.

In particular, current observations favor models in which the DE equation of state crosses the phantom divide from $w_{\rm DE} < -1$ to $w_{\rm DE} > -1$, occurring at a critical redshift $0 < z_c < 1$ \cite{DESI:2024aqx,DESI:2025fii}. Such a crossing may be realized by introducing a ghost scalar field with a negative kinetic term \cite{Caldwell:1999ew}, in addition to a quintessence field with a positive kinetic term $X$. However, it typically proceeds from $w_{\rm DE} > -1$ to $w_{\rm DE} < -1$ \cite{Feng:2004ad,Guo:2004fq}. Moreover, the presence of a ghost 
field renders the vacuum unstable due to its unboundedly negative energy density \cite{Carroll:2003st,Cline:2003gs}.

One way to achieve $w_{\rm DE} < -1$ without introducing ghosts is to extend scalar-field theories to include field derivative self-interactions or nonminimal couplings to gravity. For example, in covariant Galileon models \cite{Deffayet:2009wt}, there exists a tracker solution along which $w_{\rm DE}$ evolves from $-2$ during the matter era to the asymptotic de-Sitter value $-1$ \cite{DeFelice:2010pv}. In this case, it is possible to avoid both ghost and Laplacian instabilities, but $w_{\rm DE}$ does not enter the region $w_{\rm DE} > -1$ at low redshifts. Moreover, the tracker solution for covariant Galileons is ruled out by joint analyses of CMB, SN~Ia, and BAO data, mostly due to the large deviation of $w_{\rm DE}$ from $-1$ from the matter era to today \cite{Nesseris:2010pc,Neveu:2016gxp,Peirone:2017vcq}.

If the quadratic kinetic term $X^2$ is included in the Lagrangian of covariant Galileons, known as the Galileon Ghost Condensate (GGC) \cite{Peirone:2019aua}, the approach to the tracker is hindered by this additional contribution. As a result, the deviation of $w_{\rm DE}$ from $-1$ tends to be smaller compared to that in covariant Galileons. A statistical analysis using CMB, SN~Ia, and BAO data in 2019 showed that the GGC model is favored over the cosmological constant \cite{Peirone:2019aua}. However, 
as long as linear stability conditions 
hold in the GGC model, $w_{\rm DE}$ does not cross the phantom divide at low redshifts.

The GGC model belongs to a subclass of shift-symmetric (SS) Horndeski theories \cite{Horndeski:1974wa}, in which the Lagrangian is invariant under a constant shift of the scalar field, $\phi \to \phi + c$. Such SS scalar-tensor theories are related to vector-tensor theories with broken $U(1)$ gauge symmetry, known as generalized Proca (GP) theories \cite{Heisenberg:2014rta,Tasinato:2014eka,Allys:2015sht,BeltranJimenez:2016rff}. 
In the limit where the vector field $A_{\mu}$ is replaced by the derivative of a scalar,
$A_{\mu} \to \partial_{\mu} \phi$, the Lagrangian of GP theories reduces to that of SS Horndeski theories. The DE dynamics in GP theories was studied in Ref.~\cite{DeFelice:2016yws}, showing that the evolution of $w_{\rm DE}$ mimics that of the tracker solution for covariant Galileons. This implies that $w_{\rm DE}$ remains in the 
regime $w_{\rm DE} < -1$ from the matter era 
up to the present.

In this paper, we first show that crossing the phantom 
divide from $w_{\rm DE}< -1$ to $w_{\rm DE}> -1$ 
without encountering theoretical pathologies does 
not generally occur in shift-symmetric Horndeski and GP theories with luminal gravitational-wave (GW) speed.
This includes covariant Galileons, the GGC model, 
and DE models in GP theories. 
To demonstrate such a property, we employ the effective field theory (EFT) 
of DE, which unifies Horndeski and GP theories \cite{Aoki:2021wew,Aoki:2024ktc,Aoki:2025bmj}. 
This difficulty can be overcome by breaking the shift symmetry in scalar-tensor theories.

One such approach is to introduce a nonminimal 
coupling $F(\phi) R$ with the Ricci 
scalar $R$ \cite{Boisseau:2000pr,Perivolaropoulos:2005yv,Amendola:2007nt,Tsujikawa:2008uc,
Motohashi:2010tb,Ye:2024ywg,Wolf:2024stt,Ye:2024zpk,Pan:2025psn,Wolf:2025jed,Wang:2025znm,Adam:2025kve,SanchezLopez:2025uzw}, 
including $f(R)$ models of 
late-time cosmic acceleration \cite{Hu:2007nk,Starobinsky:2007hu,Appleby:2007vb,Tsujikawa:2007xu}. 
The nonminimal coupling mediates fifth forces between the scalar field and matter. For consistency with local gravity tests and structure formation, the deviation of $w_{\rm DE}$ from $-1$ in $f(R)$ models is so tiny that they are effectively indistinguishable from the cosmological 
constant \cite{Hu:2007nk,Brax:2008hh}. 
In the presence of the nonminimal coupling $F(\phi)R$, 
even if the scalar-matter coupling is suppressed by a screening mechanism, 
the cosmological evolution of the DE scalar $\phi$ induces a time variation of 
the gravitational coupling $G$ in overdense 
regions \cite{Babichev:2011iz,Kimura:2011dc}. 
In nonminimally coupled theories,  
the stringent Lunar Laser Ranging bounds on 
the time variation of $G$ \cite{Hofmann:2018myc} suppress deviations of $w_{\rm DE}$ from $-1$ at low redshifts \cite{Tsujikawa:2019pih}.
There have also been recent studies of interacting  
DE and dark matter (DM) models that realize 
an effective equation of state crossing the phantom divide~\cite{Chakraborty:2025syu,Khoury:2025txd,Guedezounme:2025wav}. 
It is generally nontrivial for such models to be consistent with observations unless DE itself crosses 
$w_{\rm DE} = -1$~\cite{Linder:2025zxb}.

Here, we extend the GGC model by introducing a scalar potential $V(\phi)$, without invoking a nonminimal coupling or any direct interaction between DE and DM.
The potential breaks the shift symmetry and allows the phantom-divide crossing, 
leading to an appreciable deviation of 
$w_{\rm DE}$ from $-1$. 
The new model, defined by an explicit Lagrangian, 
can realize the evolution of $w_{\rm DE}$ at low redshifts required by DESI observations.

\section{SS Horndeski and GP theories} 
\label{nogo}

We first consider SS Horndeski and GP theories with luminal GW speed,  
consistent with the GW170817 
observation \cite{LIGOScientific:2017vwq}.
The dynamics of the background and perturbations can be treated in a unified manner using the EFT of DE \cite{Aoki:2021wew,Aoki:2024ktc,Aoki:2025bmj} (see also 
\cite{Gubitosi:2012hu,Bloomfield:2012ff,Gleyzes:2013ooa,Raveri:2014cka,Lagos:2016wyv}).
On the spatially flat, isotropic cosmological background described by the line element 
${\rm d}s^2 = -\bar{N}^2(t) {\rm d}t^2 + a^2(t)\,\delta_{ij}{\rm d}x^i {\rm d}x^j$, 
we include matter fields with energy density $\rho_M$ and pressure $p_M$. 
We consider a vector-field configuration 
of the form $A_{\mu}=[A_0(t),{\bm 0}]$, 
where the temporal component $A_0$ depends 
on time $t$. 
By choosing the unitary gauge and 
introducing a gauge coupling constant $g_M$, 
a preferred timelike vector field is given by $v_{\mu}=\delta^{0}_{\mu}+g_M A_{\mu}=(1+g_M A_0, {\bm 0})$, which is proportional to a unit vector $n_{\mu}$ 
orthogonal to constant-$t$ hypersurfaces. In GP theories, we have $g_M \neq 0$, in which case the 
symmetry-breaking pattern differs from that in 
Horndeski theories ($g_M = 0$).

The EFT action describing the background and linear perturbations in SS Horndeski and GP theories is given by 
${\cal S}=\int {\rm d}^4 x \sqrt{-g}\,{\cal L}+{\cal S}_M$ \cite{Aoki:2021wew},  
where ${\cal S}_M$ is the action of matter 
minimally coupled to gravity, and 
\ba
\hspace{-0.2cm}
& &
{\cal L}=
\frac{M_*^2}{2}f \left[ {}^{(3)}R+K_{\mu \nu}K^{\mu \nu}-K^2 
\right]-\Lambda-\tilde{c}\,\tilde{g}^{00}-d\,K\,, \nonumber\\
\hspace{-0.2cm}
& &~~~~~+\frac{M_2^4}{2} 
\left( \frac{\delta \tilde{g}^{00}}
{\tilde{g}^{00}_{\rm BG}} \right)^2+\frac{\bar{M}_1^3}{2} 
\frac{\delta \tilde{g}^{00}}{\tilde{g}^{00}_{\rm BG}} \delta K
+\frac{\gamma_1}{2} F_{\mu} F^{\mu}\,,
\ea
where $M_*$ is a constant, while $f$, $\Lambda$, $\tilde{c}$, $d$, 
$M_2^4$, $\bar{M}_1^3$, and $\gamma_1$ are time-dependent functions. 
Here, ${}^{(3)}R$ is the three-dimensional Ricci scalar in the 
Arnowitt-Deser-Misner decomposition of spacetime \cite{Arnowitt:1959ah}, 
$K$ is the trace of 
the extrinsic curvature $K_{\mu \nu}$, 
and $\tilde{g}^{00}=(1+g_M A_0)^2 g^{00}$ with background value 
$\tilde{g}_{\rm BG}^{00}$ and perturbation $\delta \tilde{g}^{00}$, 
and $F_{\mu}=2n^{\nu} \nabla_{[\mu}A_{\nu]}$.
The background value of $g^{00}$ is $g_{\rm BG}^{00}=-\bar{N}^{-2}$. 
Unlike Ref.~\cite{Aoki:2025bmj}, we do not incorporate a direct coupling 
between DE and dark matter.

Varying the EFT action with respect to the background lapse $\bar{N}$ and the scale factor $a$, it follows that 
\ba
\hspace{-0.7cm}
& &
3M_*^2 fH^2=\rho_{\rm DE}
+\rho_M\,,\label{back1} \\
\hspace{-0.7cm}
& &
M_*^2 \left( 2f \dot{H}+2\dot{f}H
+3f H^2 \right)
= -p_{\rm DE}-p_M\,,
\label{back2}
\ea
where a dot denotes the derivative with respect to 
$\tilde{t}=\int \bar{N}\,{\rm d}t$, 
$H=({\rm d}a/{\rm d}\tilde{t})/a$, and 
\ba
\rho_{\rm DE} &\equiv& 
\Lambda+\tilde{c} \bar{N}^{-2}
(1+g_M A_0)^2\,,\label{rhoDE} \\
p_{\rm DE} &\equiv& 
-\Lambda+\tilde{c} \bar{N}^{-2}
(1+g_M A_0)^2+\dot{d}\,.\label{pDE}
\ea
In SS Horndeski and GP theories, the EFT functions are subject to several consistency conditions \cite{Aoki:2021wew}, i.e.,  
\ba
& &
\dot{f}=0\,, 
\label{con1} \\
& &
\dot{\Lambda}+3H \dot{d}
+\dot{\tilde{c}}\,\tilde{g}^{00}_{\rm BG}=0\,,
\label{con2} \\
& &
\dot{d}=\frac{\bar{M}_1^3 [ 3\bar{M}_1^3 \dot{H}
-2 \dot{\tilde{c}}\,(1+g_M A_0)^2 \bar{N}^{-2}]}{4M_2^4}\,.  
\label{con3} 
\ea
From Eq.~(\ref{con1}), we have 
\be
M_*^2 f \equiv \Mpl^2={\rm constant}\,,
\ee
which shows the absence of a nonminimal coupling.
In GP theories, variation of ${\cal S}$ with respect to $A_0$ yields 
$\tilde{c}\,(1+g_M A_0)=0$. 
The EFT setup based on a preferred vector field
requires $1+g_M A_0 \neq 0$, so that 
\be
\tilde{c}=0\,,\qquad ({\rm GP~theories})\,.
\label{GPc}
\ee
Since $g_M A_0=0$ in SS Horndeski theories, 
the property $\tilde{c}=0$ does not generally apply. However, combining Eqs.~(\ref{back1}) 
and (\ref{back2}) with Eq.~(\ref{con2}) gives 
$(a^3 \tilde{c}/\bar{N})^{\cdot}=0$, which integrates to 
\be
\tilde{c}=\tilde{c}_0 \bar{N} a^{-3}\,,
\qquad ({\rm SS~Horndeski~theories})\,,
\label{ssc}
\ee
with constant $\tilde{c}_0$.
Note that $\tilde{c}=0$ is a tracker solution 
found in Ref.~\cite{DeFelice:2010pv}. 
Since $\tilde{c}$ decreases to 0 in an 
expanding Universe, the solution 
finally approaches the tracker.

Taking the sum of Eqs.~(\ref{rhoDE}) and (\ref{pDE}) and 
using Eq.~(\ref{con3}), we obtain 
\be
\rho_{\rm DE}+p_{\rm DE}=12\Mpl^2 \frac{\alpha_B^2}{\alpha_K}
\dot{H}+\frac{2}{\bar{N}^2} \left( \tilde{c}
+\frac{2\dot{\tilde{c}}\,\alpha_B}{H \alpha_K} \right)
(1+g_M A_0)^2,
\label{rhopde}
\ee
where $\alpha_K \equiv 4M_2^4/(H^2 \Mpl^2)$ 
and $\alpha_B \equiv -\bar{M}_1^3/(2H \Mpl^2)$. 

In GP theories, the last term in Eq.~(\ref{rhopde}) vanishes 
since $\tilde{c}=0$.
A consistent background solution corresponds to a DE equation 
of state $w_{\rm DE} \equiv p_{\rm DE}/\rho_{\rm DE} < -1$, 
with $\rho_{\rm DE} > 0$, $\dot{H} < 0$, 
and $\alpha_K > 0$ \cite{DeFelice:2016yws}.
The ghost-free condition for scalar perturbations in the small-scale 
limit requires $Q_s=\alpha_K + 6\alpha_B^2 > 0$ \cite{Aoki:2021wew,Aoki:2025bmj}. 
Provided that $\alpha_K > 0$, 
the theory remains ghost-free in the regime $w_{\rm DE} < -1$. 
To realize a phantom-divide crossing from $w_{\rm DE} < -1$ to $w_{\rm DE} > -1$ with $\rho_{\rm DE} > 0$ and $\dot{H} < 0$,  $\alpha_K$ must cross zero.
However, this would cause 
$\rho_{\rm DE} + p_{\rm DE}$ 
to diverge for a nonzero value of $\alpha_B$.
If $\alpha_B$ crosses 0 simultaneously with $\alpha_K$, it is possible to 
satisfy Eq.~(\ref{rhopde}) for 
finite values of $\rho_{\rm DE} + p_{\rm DE}$ and $\dot{H}$. 
In this case, however, $Q_s$ crosses 0, indicating the strong coupling problem 
of scalar perturbations. 
These arguments illustrate the difficulty of crossing $w_{\rm DE} = -1$ in GP theories.

In SS Horndeski theories, $\tilde{c}$ 
decreases to 0 according to Eq.~(\ref{ssc}), 
so the late-time cosmological evolution 
resembles the solution in GP theories. 
To realize the behavior $w_{\rm DE} < -1$ 
in the past, the solution needs to approach
the tracker to some extent at least by redshifts $z > {\cal O}(1)$, 
after which the last term of Eq.~(\ref{rhopde}) can be negligible.
Then, provided $\rho_{\rm DE} > 0$ and 
$\dot{H} < 0$, we have $w_{\rm DE} < -1$ for $\alpha_K > 0$. Crossing $w_{\rm DE} = -1$ 
at low redshifts would again require a sign 
change of $\alpha_K$.  
The ghost-free condition in Horndeski theories 
is given by \cite{Aoki:2021wew,Aoki:2025bmj} 
\be
Q_s = \alpha_K+6\alpha_B^2+6\Omega_{\tilde{c}}>0\,,
\label{Qs}
\ee
where $\Omega_{\tilde{c}} \equiv \tilde{c} \dot{\phi}^2/(3H^2 \Mpl^2)$. 
Using the dictionary between EFT functions and Horndeski couplings $G_{2,3}(X)$ \cite{Aoki:2021wew,Aoki:2025bmj}, both $\alpha_K$ and $\alpha_B$ vanish at $\dot{\phi}=0$ for $G_{2,3}(X)$ involving power-law functions $X^p$ with integer $p \geq 0$.
Since $\Omega_{\tilde{c}}$ also vanishes at 
$\dot{\phi}=0$, the crossing $\alpha_K=0$ 
implies $Q_s=0$, signaling strong coupling. 
Thus, in both GP and SS Horndeski theories, realizing 
the phantom-divide crossing suggested by DESI data is difficult without encountering theoretical pathologies.

\section{Models with the phantom divide crossing} 
\label{dividesec}

As discussed in Sec.~\ref{nogo}, realizing the phantom-divide crossing requires breaking the shift symmetry, 
which can be achieved by 
introducing a scalar potential $V(\phi)$. We consider 
the four-dimensional Lagrangian
\be
{\cal L} = \frac{\Mpl^2}{2}R + a_1 X + a_2 X^2 
+ 3 a_3 X \square \phi - V(\phi)\,,
\label{Lcon}
\ee
with $X = -\partial_\mu \phi \, \partial^\mu \phi / 2$, where
$a_1$, $a_2$, and $a_3$ are constants. 
The SS GGC model \cite{Peirone:2019aua} 
corresponds to the limit $V(\phi) \to 0$. 
The Lagrangian~(\ref{Lcon}) involves 
no nonminimal coupling, so that $\dot f = 0$. 
Since the shift symmetry is broken, 
the properties~(\ref{ssc}) and~(\ref{rhopde}) no longer hold. 

The constants $a_2$ and $a_3$ can be expressed as 
$a_2 = \tilde{a}_2 / \Lambda_2^4$ and 
$a_3 = \tilde{a}_3 / \Lambda_3^3$,
where $\tilde{a}_2$ and $\tilde{a}_3$ are 
dimensionless constants at most of order 1, 
and the mass scales $\Lambda_2$ and $\Lambda_3$ 
relevant to today's cosmic 
acceleration are $\Lambda_2 
= \sqrt{M_{\rm pl} H_0}$ and $\Lambda_3 = (M_{\rm pl} H_0^2)^{1/3}$, 
with $\Mpl \simeq 10^{18}~\mathrm{GeV}$ and
$H_0 \simeq 10^{-42}~\mathrm{GeV}$.
Our model should be regarded as a low-energy EFT 
valid below the scale 
$\Lambda_3 \simeq 10^{-22}~\mathrm{GeV} \simeq 10^{20} H_0$, above which the theory becomes strongly coupled 
due to the dominance of higher-derivative operators required for its ultraviolet (UV) completion~\cite{Luty:2003vm,deRham:2011qq,deRham:2018red}.

In the low-energy EFT regime of theories with 
weakly broken Galileon
invariance \cite{Pirtskhalava:2015nla} (including our model), quantum loop corrections arising from a vertex involving 
the scalar derivatives $X^n$ and a single graviton 
are estimated to be of the order 
${\cal L}_c=\Lambda_3^4 [X/(\Mpl \Lambda_3^3)]^n 
=\Lambda_3^4 [X/(\Mpl^2 H_0^2)]^n$, 
where $n$ is a positive integer. 
Since $X$ is at most of the order of today's critical density, 
$\rho_0 = 3 M_{\rm pl}^2 H_0^2$, loop corrections 
are highly suppressed:
${\cal L}_c \lesssim \Lambda_3^4 \simeq (H_0/\Mpl)^{2/3}
\rho_0 \ll \rho_0$. 
Compared to the term $\tilde{a}_2 X^2/\Lambda_2^4$, loop-generated derivative operators of order higher than $X^2$ 
are also strongly suppressed due to the mass hierarchy $\Lambda_2 \gg \Lambda_3$~\cite{Pirtskhalava:2015nla}.

The potential $V(\phi)$ also receives quantum corrections 
from its self-coupling, resulting in the effective potential
$V_{\rm eff}(\phi)=\exp [\pm \mu^2/(32\pi^2)
\,{\rm d}^2/{\rm d}\phi^2]V(\phi)$, where $\mu$ 
is a UV cutoff scale \cite{Garny:2006wc} (see also Ref.~\cite{Doran:2002bc}).
The linear potential $V(\phi) = m^3 \phi$, which restores 
Galileon symmetry in the Minkowski limit with 
$\Lambda_2 \to \infty$ \cite{Nicolis:2008in}, 
does not receive such quantum corrections. 
Indeed, the model with 
$a_2 = 0$ and $V(\phi) = m^3 \phi$ represents 
the minimal theoretical 
setup that allows for the phantom-divide crossing.
Beyond allowing an extension to $a_2 \neq 0$, 
Galileon theory can 
also be generalized to include a broader class 
of potentials that receive only small quantum corrections. 
One example is provided by the exponential potential \cite{Wetterich:1987fm,Ratra:1987rm,Wetterich:1994bg,Copeland:1997et,Gasperini:2001pc}:
\be
V(\phi)=V_0 e^{-\lambda \phi/\Mpl}\,,
\label{exp}
\ee
where the constant $\lambda$ is at most of order 1. 
Quantum corrections modify $V_0$ to 
$\tilde{V}_0=V_0 \exp [\pm \lambda^2 \mu^2/(32 \pi^2 \Mpl^2)]$, so that $\tilde{V}_0$ remains very close 
to $V_0$ for $\mu \approx \Lambda_3 \ll M_{\rm pl}$.

We study the background dynamics for the exponential 
potential (\ref{exp}). We introduce the dimensionless variables
$x_1 \equiv \dot{\phi}/(\sqrt{6} \Mpl H)$, 
$x_2 \equiv a_2 \dot{\phi}^4/(4 \Mpl^2 H^2)$, 
$x_3 \equiv -3a_3 \dot{\phi}^3/(\Mpl^2 H)$, and 
$x_4 \equiv V(\phi)/(3\Mpl^2 H^2)$. 
In the matter sector, we include nonrelativistic matter 
(energy density $\rho_m$, 
negligible pressure) and radiation (energy density $\rho_r$, pressure $p_r = \rho_r/3$).
Then, the background equations of motion are of the 
forms (\ref{back1}) and (\ref{back2}), with $M_*^2 f=\Mpl^2$, 
$\rho_M=\rho_m+\rho_r$, $p_M=\rho_r/3$, and 
\ba
\rho_{\rm DE} &=& 3\Mpl^2 H^2 \left( a_1 x_1^2+
x_2+x_3+x_4 \right)\,,\\
p_{\rm DE} &=& \Mpl^2 H^2 
\left( 3 a_1 x_1^2 +x_2 -x_3 \epsilon_{\phi} 
-3x_4 \right)\,,
\ea
where $\epsilon_{\phi} \equiv \ddot{\phi}/(H \dot{\phi})$. 
Then, the DE equation of state is given by 
$w_{\rm DE}=p_{\rm DE}/\rho_{\rm DE}$.  
The dimensionless variables $x_{1,2,3,4}$
and the radiation density parameter 
$\Omega_r = \rho_r/(3 \Mpl^2 H^2)$ satisfy
\ba
\hspace{-0.7cm}
& &
x_1'= x_1 \left( \epsilon_{\phi} - h \right)\,,\qquad 
x_2'=2x_2 \left( 2\epsilon_{\phi} - h \right)\,,\nonumber \\
\hspace{-0.7cm}
& &
x_3'=x_3 \left( 3\epsilon_{\phi} - h \right)\,,\quad~\,
x_4'= -x_4 ( \sqrt{6} \lambda x_1 + 2h )\,,\nonumber \\
\hspace{-0.7cm}
& &
\Omega_r'=-2\Omega_r(2+h)\,,
\ea
where $h \equiv \dot{H}/H^2$, 
$\lambda \equiv -(\Mpl/V)({\rm d}V/{\rm d}\phi)$, and 
a prime represents the derivative with respect to $\ln a$. 
Differentiating Eq.~(\ref{back1}) with respect to $t$ and 
combining it with Eq.~(\ref{back2}) yields expressions 
that determine $\epsilon_{\phi}$ and $h$. 
From Eq.~(\ref{back1}), the matter density parameter 
$\Omega_m=\rho_m/(3\Mpl^2 H^2)$ is expressed as
\be
\Omega_m=1-a_1 x_1^2-x_2-x_3-x_4-\Omega_r\,.
\ee

Since the Lagrangian (\ref{Lcon}) belongs to a subclass of 
Horndeski theories with a luminal 
gravitational-wave speed \cite{Kobayashi:2011nu}, 
the second-order action of tensor perturbations is the same as 
that in General Relativity (GR).
The ghost-free condition for the scalar perturbation is 
\be
Q_s=(3/2)( 4 a_1 x_1^2 + 8 x_2 + 4x_3 + x_3^2)>0\,.
\ee
The Laplacian instability of scalar perturbations is absent if
\be
c_s^2=\frac{(2+6\epsilon_{\phi})x_3-x_3^2
-4h-6\Omega_m-8\Omega_r}
{3( 4 a_1 x_1^2 + 8 x_2 + 4x_3 + x_3^2)}>0\,.
\ee

Provided that the inequality $\{ |a_1 x_1^2|, |x_2|, |x_4|\} \ll |x_3| \ll 1$ 
holds during the early epoch, we have 
$\epsilon_{\phi} \simeq (\Omega_r-3)/4$, $h \simeq -(3+\Omega_r)/2$, 
$w_{\rm DE} \simeq (3-\Omega_r)/12$, $Q_s \simeq 6x_3$, and 
$c_s^2 \simeq (5+\Omega_r)/12$, so that the ghost is absent for 
$x_3>0$. Introducing the quantity $y \equiv x_3/(a_1 x_1^2)$ and 
taking the limits $x_2 \to 0$ and $x_4 \to 0$, we obtain 
\be
y'=-\frac{y(y+2)[9+\Omega_r+3(y+1)a_1x_1^2]}
{4(y+1)+y^2 a_1 x_1^2}\,.
\ee
This allows the existence of the fixed point $y=-2$, i.e., 
$x_3=-2a_1 x_1^2$, which corresponds to the tracker 
present for covariant Galileons \cite{DeFelice:2010pv}.
Under the condition $x_3>0$, the tracker exists for $a_1<0$. 
For the initial condition $|a_1| x_1^2 \ll x_3$, $y$ is initially 
in the range $|y| \gg 1$ and can approach $y=-2$ 
as $x_1^2$ grows faster than $x_3$.
Along the tracker, we have approximately  
$w_{\rm DE} \simeq -(6+\Omega_r)/[3(1-a_1 x_1^2)]$, 
$Q_s \simeq -6 a_1 x_1^2 (1-a_1 x_1^2)$, and 
$c_s^2 \simeq [4+\Omega_r+a_1 x_1^2 (3-a_1 x_1^2)]
/[3(1-a_1 x_1^2)^2]$. The future de Sitter fixed point 
corresponds to $a_1 x_1^2=-1$, $x_3=2$, $\Omega_m=0$, and 
$\Omega_r=0$. Provided $a_1<0$ and $a_1 x_1^2$ 
decreases toward $-1$, $w_{\rm DE}$ evolves from $-2$ 
(matter era) to $-1$ (de Sitter era) without crossing the 
phantom divide, while satisfying the stability 
conditions $Q_s>0$ and $c_s^2>0$.

\begin{figure}[ht]
\begin{center}
\includegraphics[height=2.7in,width=3.3in]{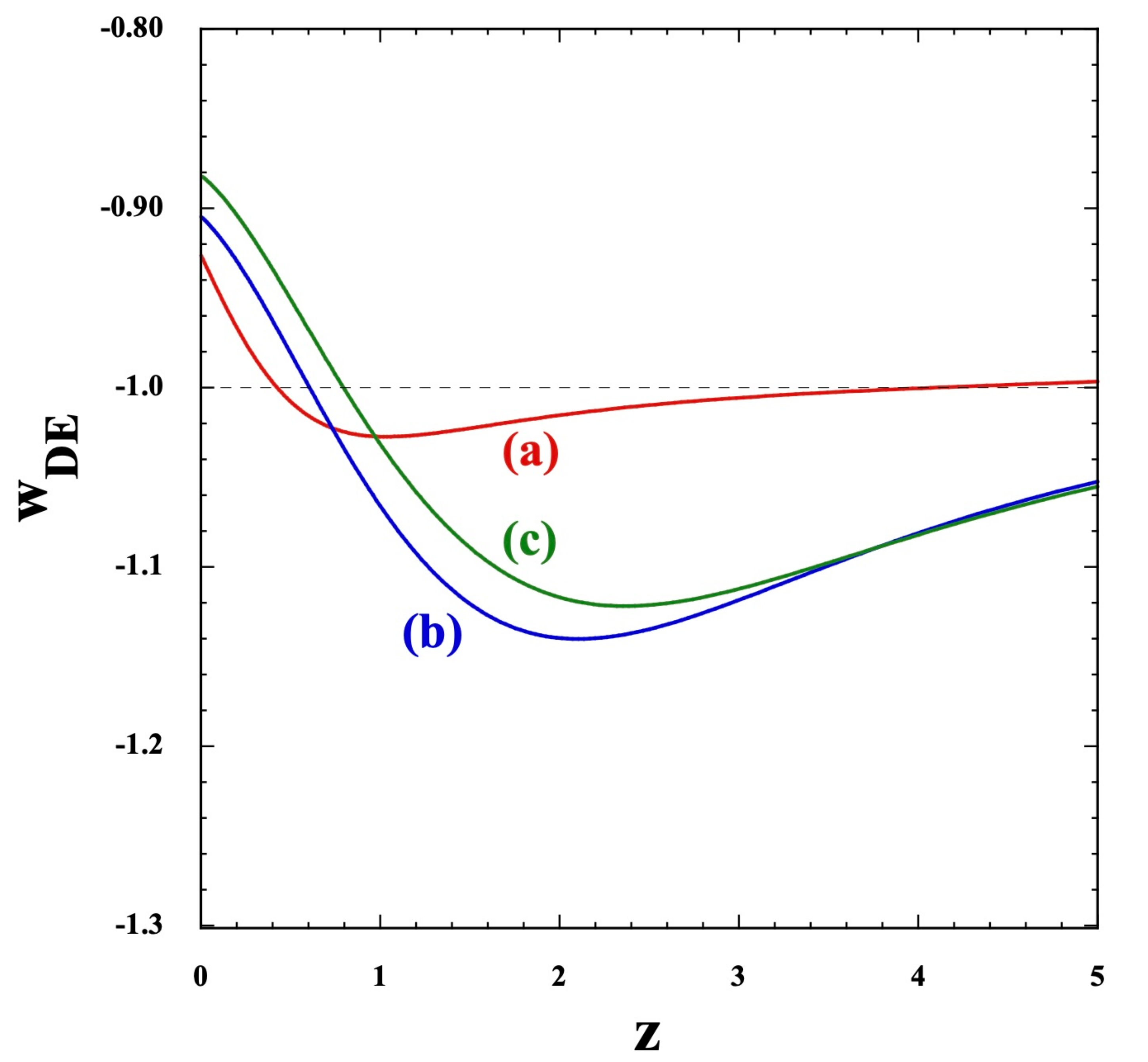}
\end{center}
\caption{Evolution of $w_{\rm DE}$ versus the redshift 
$z$ for $a_1=-1$, $\lambda=1$, and three different cases: 
(a) $x_1=0.442$, $x_2=0$, $x_3=0.519$, 
$x_4=0.356$, 
(b) $x_1=0.792$, $x_2=0.512$, $x_3=0.672$, 
$x_4=0.124$, and 
(c) $x_1=0.792$, $x_2=0.569$, $x_3=0.605$, 
$x_4=0.134$ at $z=0$, respectively, where 
$\Omega_{m}=0.32$ and 
$\Omega_r=9.0 \times 10^{-5}$ today.
Case (a) corresponds to $a_2 = 0$, 
whereas $a_2 \neq 0$ in cases (b) and (c).
\label{fig1}}
\end{figure}

The above cosmological evolution of covariant Galileons 
is modified when $x_2 \neq 0$ and/or $x_4 \neq 0$. 
Let us first consider the case $x_2 = 0$ and $x_4 \neq 0$. 
In the past, the solution must partially approach the tracker 
to enter the region $w_{\rm DE} < -1$. 
To achieve a transition to $w_{\rm DE} > -1$ at low redshifts, 
we require that the potential $V(\phi)$ is the dominant source 
of today's cosmic acceleration.
In case (a) of Fig.~\ref{fig1}, we show an example of the  
phantom-divide crossing, which occurs at $z_c = 0.42$. 
As today's value of $x_1$ decreases for given $x_3$ and $x_4$ 
at $z=0$, the transition redshift $z_c$ tends to increase, and 
eventually $w_{\rm DE}$ remains in the region 
$w_{\rm DE} \geq -1$ throughout the evolution. 
For increasing $x_1$, the deviation of $w_{\rm DE}$ from $-1$ in the region 
$w_{\rm DE} < -1$ tends to become larger, while the transition redshift 
decreases and can even reach $z_c < 0$.
If we require $z_c$ to lie in the range $0 < z_c < 1$ for models with 
$x_2 = 0$ and $x_4 \neq 0$, a significant deviation of $w_{\rm DE}$ from 
$-1$ typically does not occur.

For models with $x_2 \neq 0$ and $x_4 \neq 0$, the presence 
of the term $a_2 X^2$ generally allows a larger deviation of $w_{\rm DE}$ from $-1$. 
In case (b) of Fig.~\ref{fig1}, the transition from $w_{\rm DE} < -1$ to 
$w_{\rm DE} > -1$ occurs around $z_c = 0.61$, with 
a more significant deviation of $w_{\rm DE}$ from $-1$ 
in the past compared to case (a).
For larger today's values of $x_2$, the transition redshift tends to increase, 
e.g., $z_c = 0.80$ in case (c). 
This behavior is attributed to the fact that the approach to the tracker 
is halted earlier.
We note that there is a de Sitter fixed point characterized by 
$x_2 = -3 a_1 x_1^2 - 3$, $x_3 = 2 a_1 x_1^2 + 4$, 
$x_4 = 0$, and $\Omega_r = 0$. 
In the future, the potential ceases to 
contribute to the dynamics, so that $w_{\rm DE}$ approaches 
$-1$ as in the GGC model \cite{Peirone:2019aua}. 
In the asymptotic past, the terms $x_2$ and $x_4$ are subdominant 
to $x_3$, and hence the dynamics is similar to that 
of covariant Galileons discussed above. 
In all cases shown in Fig.~\ref{fig1}, we confirmed that 
the stability conditions $Q_s > 0$ and $c_s^2 > 0$ are satisfied throughout the cosmological evolution.

Let us also discuss the evolution of linear scalar 
perturbations, with the perturbed line element
\be
{\rm d}s^2=-\left( 1+2\Psi \right) {\rm d}t^2
+a^2(t) \left(1-2\Phi \right) \delta_{ij} {\rm d}x^i 
{\rm d}x^j\,,
\ee
where $\Psi$ and $\Phi$ are the gravitational potentials.
In Fourier space, we relate $\Psi$ and 
$\Psi + \Phi$ to the total 
matter density contrast $\delta = \sum_{i} \rho_i \delta_i / \rho_M$ (with $i = m, r$) as 
\ba 
\label{mudef}
k^2\Psi &=& -4\pi G_{\rm N} a^2\mu(a,k)
\rho_M \delta\,, \\
k^2(\Psi+\Phi) &=& -8\pi G_{\rm N} a^2 \Sigma(a,k)
\rho_M \delta\,,
\ea
where $k$ is the comoving wavenumber, 
$\delta$ is the density 
contrast, and $G_{\rm N}=(8\pi M_{\rm pl}^2)^{-1}$ is the Newtonian gravitational constant.
The dimensionless quantities $\mu$ and $\Sigma$ 
characterize the effective gravitational couplings for matter and light, 
respectively \cite{Amendola:2007rr,Bertschinger:2008zb,Zhao:2008bn}.

\begin{figure}[ht]
\begin{center}
\includegraphics[height=2.6in,width=3.3in]{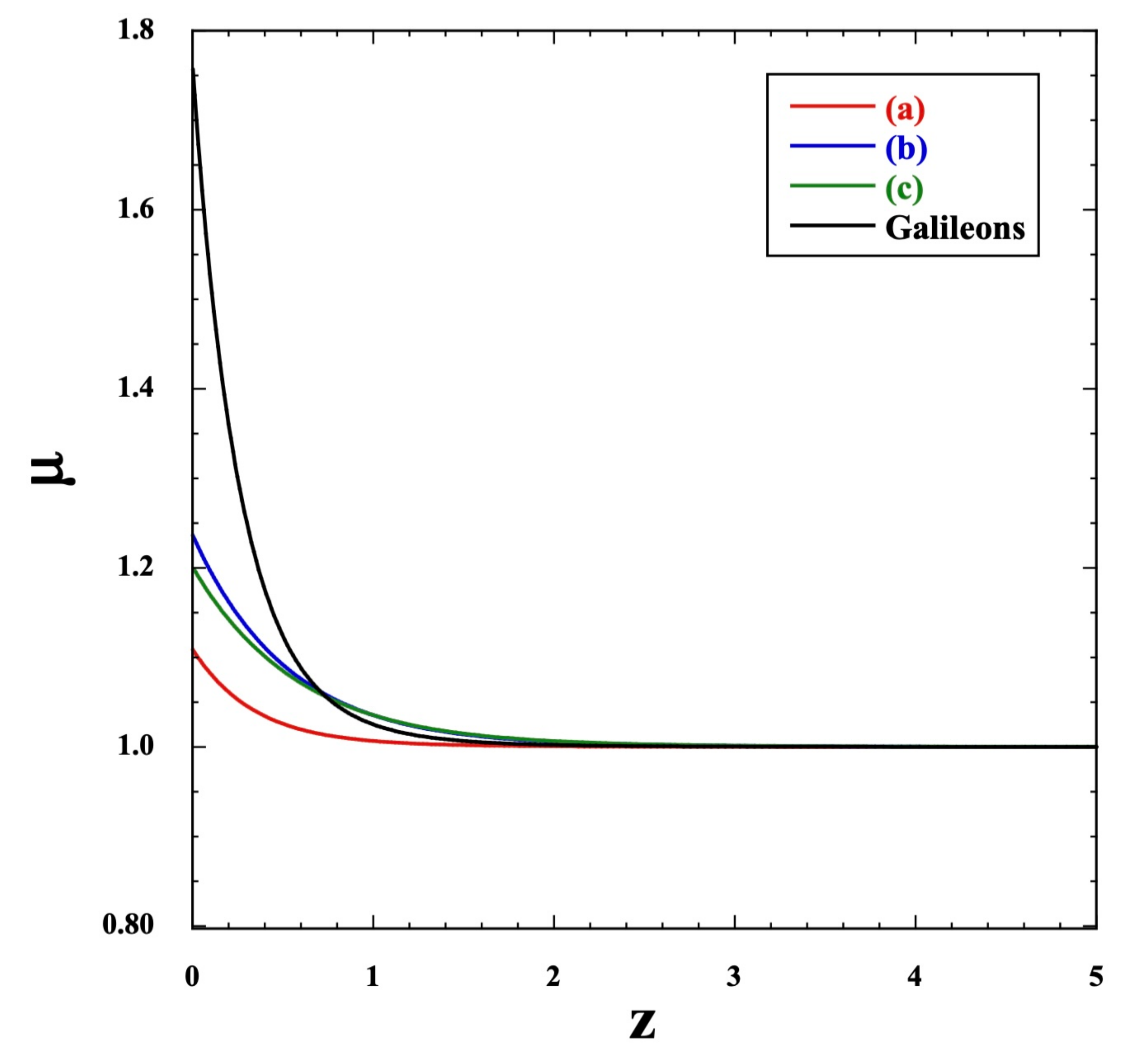}
\end{center}
\caption{Evolution of $\mu$ versus redshift $z$ for cases (a)-(c) 
in Fig.~\ref{fig1}, and for the tracker solution of 
covariant Galileons ($x_2=0$, $x_4=0$).
\label{fig2}}
\end{figure}

Under a quasi-static approximation for modes deep inside 
the sound horizon, the density contrast 
obeys \cite{DeFelice:2011hq,Kase:2018aps}
\be
\ddot{\delta}+2H \dot{\delta}-4 \pi G_{\rm N} \mu 
\rho \delta=0\,,
\label{delta}
\ee
where
\be
\mu=\Sigma=1+\frac{x_3^2}{2Q_s c_s^2}\,.
\label{mu}
\ee
In the absence of ghost and Laplacian instabilities, the Galileon term enhances gravitational interactions.
The tracker solution of covariant Galileons obeys 
$x_3 = -2a_1 x_1^2$, so that $x_3$ increases toward the 
de Sitter value $x_3 = 2$ (with $a_1 x_1^2 = -1$). 
As seen in Fig.~\ref{fig2}, this leads to a large 
deviation of $\mu$ from 1 at the present epoch, 
a behavior disfavored by observational data \cite{Peirone:2017vcq}.
For cases (a), (b), and (c), the solutions do not fully 
reach the tracker by today, with the potential dominating 
the energy density at low redshifts. 
This results in smaller present-day values of $x_3$ than in the covariant 
Galileon case, thereby suppressing the deviation of $\mu$ from 1. 
Thus, our model allows the possibility of being compatible with measurements 
of the cosmic growth rate.

\begin{figure}[ht]
\begin{center}
\includegraphics[height=2.6in,width=3.3in]{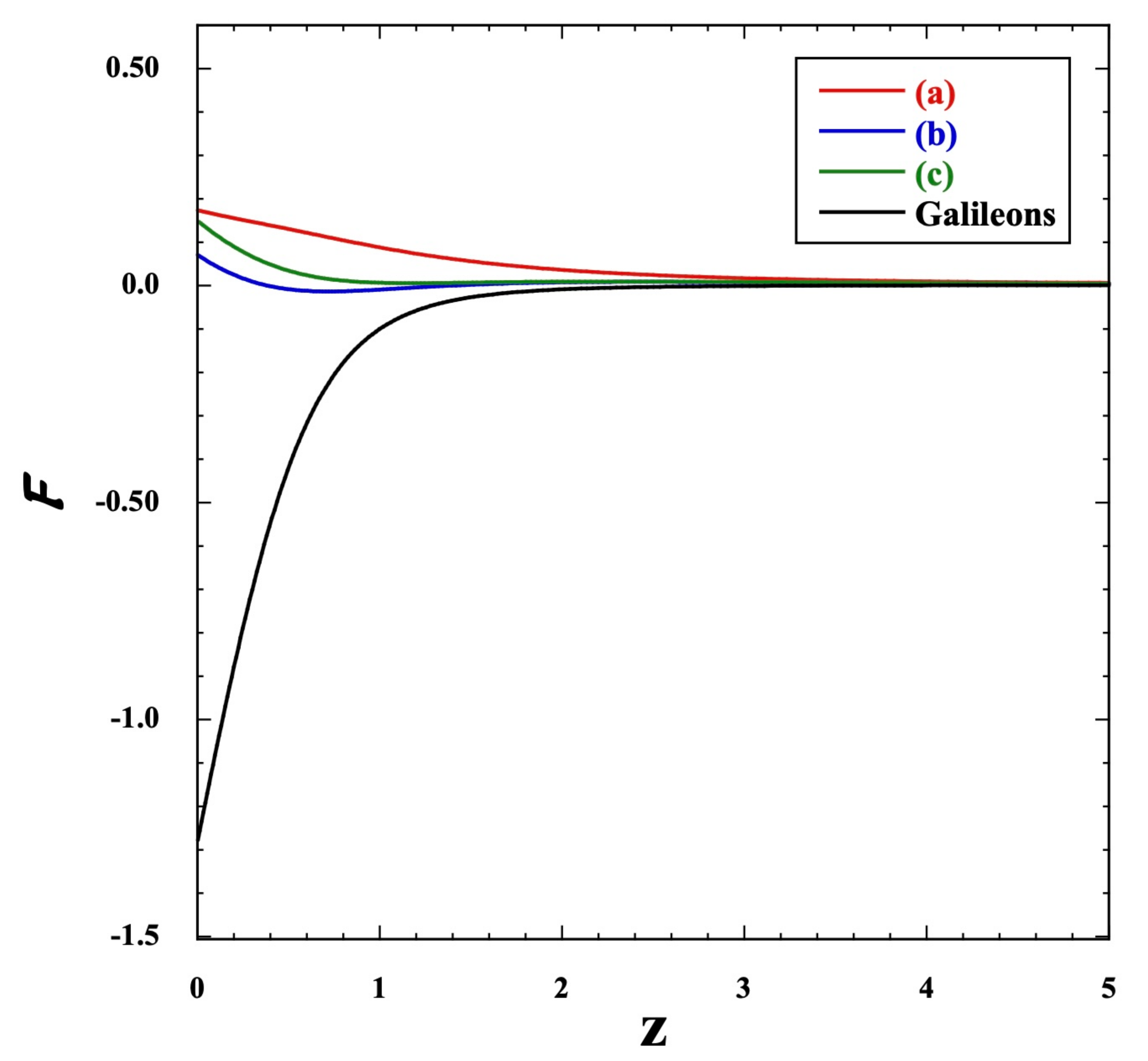}
\end{center}
\caption{Evolution of ${\cal F}$ versus redshift $z$ for 
cases (a)-(c) in Fig.~\ref{fig1}, together with that 
for covariant Galileons.
\label{fig3}}
\end{figure}

For covariant Galileons, it is known that the cross-correlations between the Integrated Sachs-Wolfe (ISW) effect in the CMB and galaxy fluctuations \cite{Boughn:2001zs} tend 
to be negative \cite{Kimura:2011td,Renk:2017rzu,Kable:2021yws}, 
a property that is disfavored by observational 
data \cite{Giannantonio:2008zi,Stolzner:2017ged}.
A quantity characterizing the sign of the 
cross-correlations is defined as \cite{Kable:2021yws}
\be
{\cal F} \equiv 1-D'/D-\Sigma'/\Sigma\,,
\ee
where $D$ is the growth function of $\delta$, 
and $\Sigma$ is given in Eq.~(\ref{mu}). 
We numerically solve Eq.~(\ref{delta}) from the deep matter era 
using the same initial conditions as in GR ($\delta' = \delta$), and 
compute the function ${\cal F}$. 
As seen in Fig.~\ref{fig3}, the covariant Galileon produces a 
strongly negative ${\cal F}$ at low redshifts, leading to negative ISW-galaxy cross-correlations.
In case (a), however, ${\cal F}$ is always positive 
throughout the evolution. 
In case (b), ${\cal F}$ is temporally negative 
for $0.39<z<1.40$, but it changes to positive for $z<0.39$. Since the cross-correlation function $C_l^{\rm Tg}$ is 
an integral from $z=0$ to the recombination epoch ($z \simeq 1090$) involving ${\cal F}$, temporarily entering 
the region ${\cal F} < 0$ does not necessarily imply $C_l^{\rm Tg} < 0$.
In case (c), ${\cal F}$ is always positive and 
hence $C_l^{\rm Tg}>0$. 
These results suggest that our model can be compatible with observations 
of ISW-galaxy cross-correlations.

\section{Conclusions} 
\label{consec}

We have shown that breaking the shift symmetry in Horndeski theories is crucial for realizing a crossing of $w_{\rm DE} = -1$ at low redshifts.
If future observations, including those from the Euclid satellite \cite{Amendola:2016saw}, were to firmly confirm the phantom-divide crossing, both SS Horndeski and GP theories could be ruled out as viable DE models.

We proposed an explicit model for the phantom-divide crossing, described by the 
Lagrangian (\ref{Lcon}), in which the potential $V(\phi)$ is incorporated into the GGC model to break the shift symmetry. 
As seen in case (a) of Fig.~\ref{fig1}, even the model with $a_2 = 0$ allows a crossing of $w_{\rm DE} = -1$, albeit with a moderate deviation of $w_{\rm DE}$ from $-1$. 
In cases (b) and (c) of Fig.~\ref{fig1}, corresponding to models with $a_2 \neq 0$, the crossing of $w_{\rm DE} = -1$ can occur in the redshift range 
$0 < z_c < 1$, with larger deviations of $w_{\rm DE}$ from $-1$ than in case (a).
We also showed that the models in cases (a)-(c) exhibit suppressed cosmic growth rates at low redshifts compared to the tracker solution of covariant Galileons, while allowing the possibility of positive ISW-galaxy 
cross-correlations.

Since our model has neither nonminimal couplings nor direct couplings of DE to baryons,  
we do not need to worry about fifth-force propagation in overdense regions.
Unlike nonminimally coupled DE models, 
such as those in $f(R)$ gravity \cite{Hu:2007nk,Brax:2008hh}, 
this framework permits larger deviations of 
$w_{\rm DE}$ from $-1$ both before and after the phantom-divide crossing.
It is therefore interesting to constrain the model using 
DESI together with other data \cite{comment}, 
and with upcoming Euclid data. 

\section*{Acknowledgements}

The author thanks Antonio De Felice for valuable discussions. 
This work was supported by JSPS KAKENHI Grant No.~22K03642 and 
Waseda University Special Research Projects (Nos.~2025C-488 and 2025R-028).

\bibliographystyle{mybibstyle}
\bibliography{bib}

\end{document}